# Multi-bunch injection for SSRF storage ring

Jiang Bocheng, Lin Guoqiang, Wang Baoliang, Zhang Manzhou, Yin Chongxian, Yan Yingbing, Tian Shunqiang, Wang Kun
Shanghai Institute of Applied Physics, Chinese Academy of Sciences, Shanghai 201800, China

*Abstract*

The multi-bunch injection has been adopt at SSRF which greatly increases the injection rate and reduces injection time compared to the single bunch injection. The multi-bunch injection will massively reduce the beam failure time during users' operation and prolong pulsed injection hardware lifetime. In this paper, the scheme to produce multi-bunches for the RF electron gun is described. The refilling result and the beam orbit stability for top up operation is discussed.

## 1、Introduction

The third generation light sources are the working horses for X-ray science worldwide. The beam availability time is usually more than 98% for advanced light sources. Great efforts have been made to increase mean time between failures (MTBF) and to decrease mean down time (MDT) to satisfy users' requirement. One of the potential method to decrease MDT is to increase the injection rate as the motivation of this paper.

The injection statuses for some of the light sources are listed below. Most of them using single bunch injection scheme. The advantage of single bunch injection is that bunch charge can be kept highly uniform for top up operation which will benefit beam orbit stability, cause the beam position monitor (BPM) is sensitive to the filling pattern when approaching orbit stability to sub-micron level[1].

Table 1. Injection status of 3$^{rd}$ generation light sources[2~8]

| Synchrotron | Topup bunch injection mode | Booster repetition rate |
|---|---|---|
| Diamond | single bunch | 5Hz |
| Soleil | single bunch | 3Hz |
| Spear 3 | single bunch | 10Hz |
| ALS | multi bunches | 1Hz |
| SLS | single bunch | 3Hz |
| Spring8 | single or trains | 1Hz |
| APS | single bunch | 2Hz |

SSRF is an advanced 3$^{rd}$ generation light source in main land China. The storage ring circumference is 432m, beam energy 3.5GeV, holding a natural emittance 3.9nm·rad.

The injector of SSRF storage ring includes a 150 MeV linac and a 3.5 GeV booster. The booster operate at a repetition rate 2 Hz. For single bunch injection, filling the storage ring from 0 mA to 240mA usually takes around 20 minutes. The limitations are from beam charge emitted by the cathode of the RF gun and from the booster repetition rate. For the top up operation, refilling 1 mA takes 15 seconds, injecting 30 tiny single bunches to where bunches' charge are lowest in the ring, keeping the filling pattern uniformly. In 2014 SSRF started investigating multi-bunch injection scheme to reduce the injection time. In the following sections details of the study is presented.

## 2、Multi bunches generation for the linac

To generate multi bunches, the pulsed voltage of grid cap has been stretched from 3ns to 12ns using reflecting lines. The method is pretty simple and economy. The measured voltage pulse is as shown in Fig. 1. The stretched pulse will create a 12 ns long electron bunch from the electron gun. The long bunch is then sent to a 500MHz sub harmonic buncher. 5~6 bunches is then formed at the exit of the buncher with bunch length 300ps, energy 10MeV. Bunches are accelerated to 157MeV by the linac through 4 travelling wave tubes.

The energy divergence of the bunches are 0.74% as shown in Fig. 2, which is worse than single bunch mode 0.5%. As energy acceptance for the booster injection is 2.0% which is bigger than the energy divergence of the multi bunches which ensure a large injection efficiency for the booster.

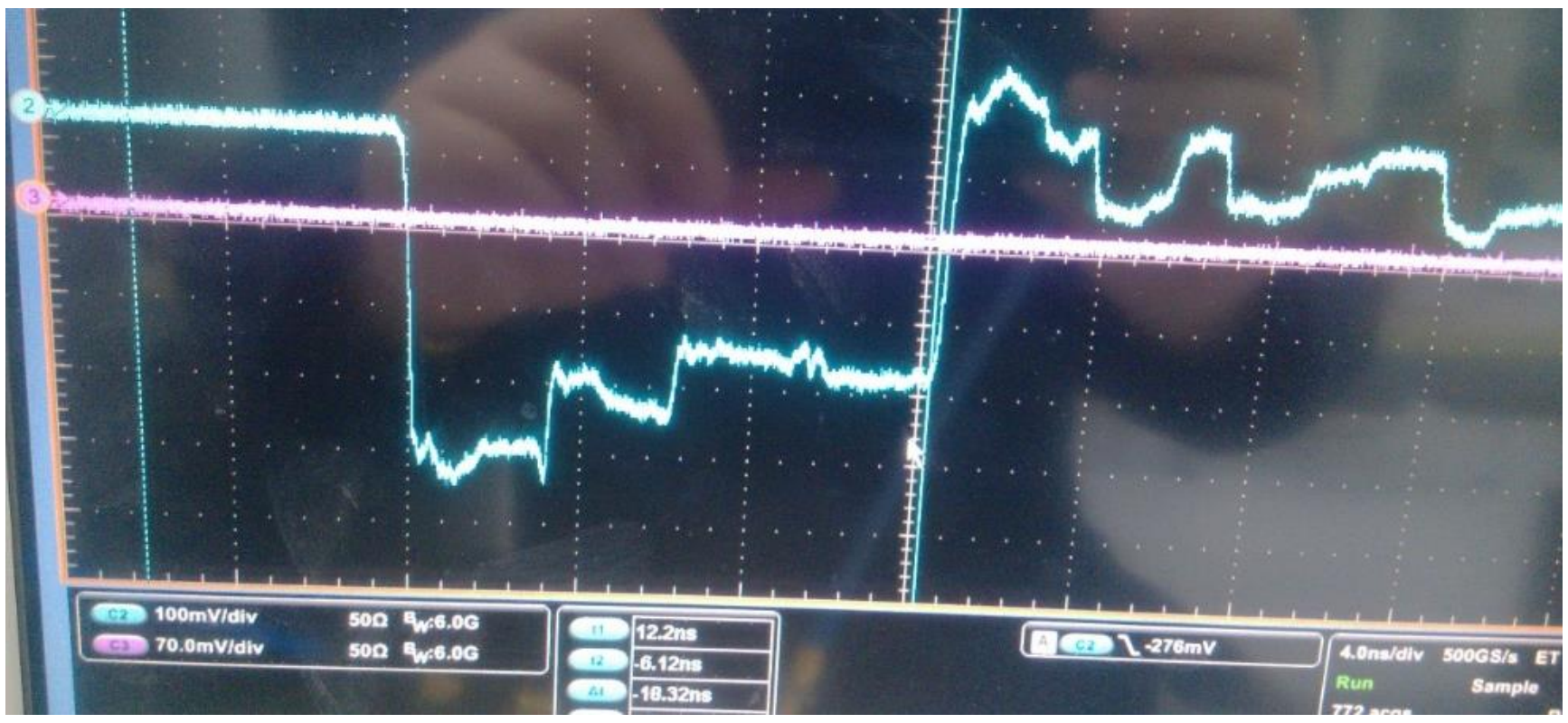

Figure 1. Measured voltage pulse on the grid cap. （The voltage after reflection gets decreased）

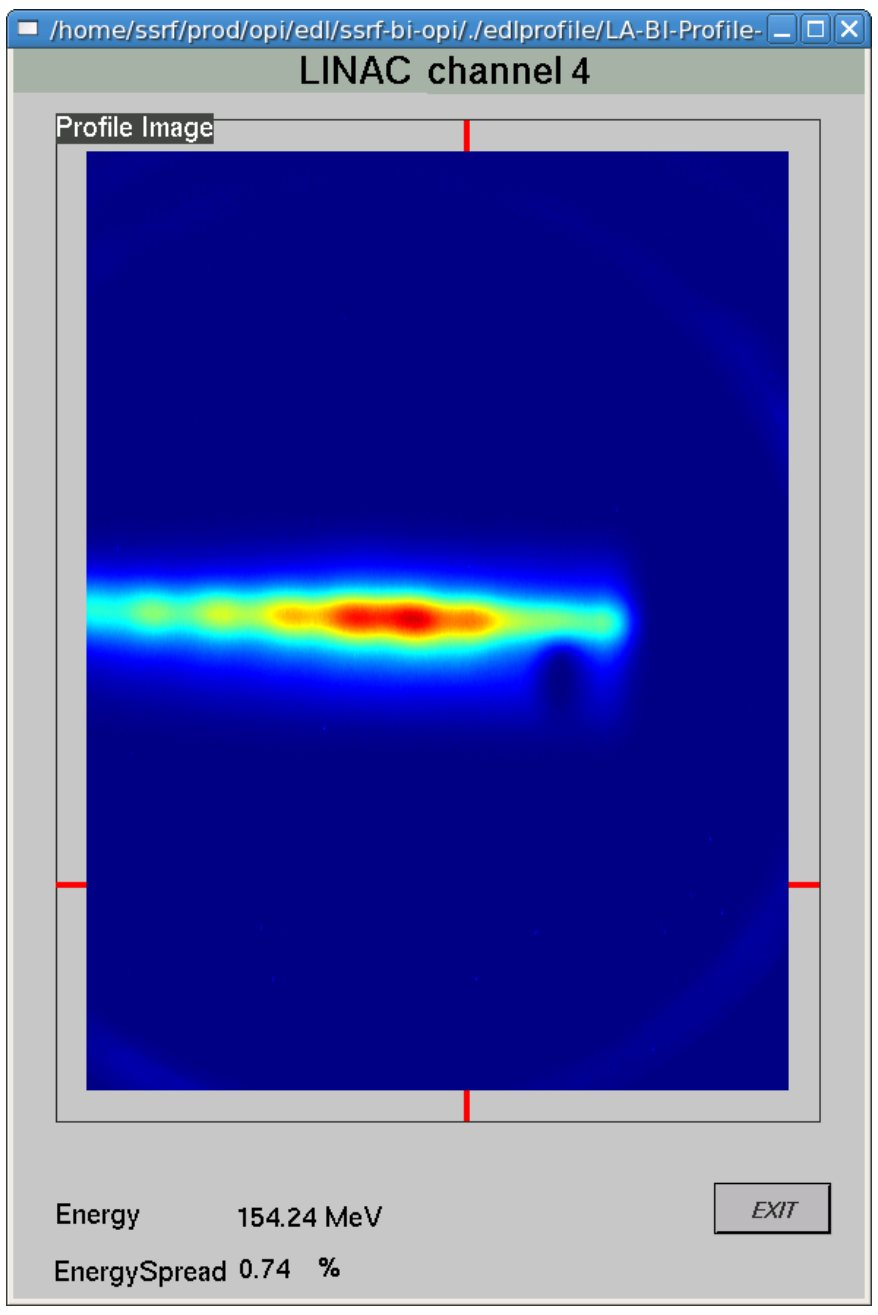

Figure 2. Measured energy spread at the end of the linac for multi bunches

## 3、Top up injection efficiency and bunch charge uniformity

For top up operation, injection efficiency from booster to storage ring is a critical parameter for radiation dose concern. The injection efficiency interlock threshold for SSRF is set to 50%. When injection efficiency drops below 50%, the top up injection will be stopped and the alarm will ring.

For multi-bunch injection the situation is not much different from single bunch injection. The energy deviation among multi bunches from linac will be fully damped during ramping of energy by the booster. The quality of extracted bunches from the booster is fully depended on the booster conditions. The pulses of the booster extraction kickers and of the storage ring injection kickers have large flat tops of around 125 ns. The multi bunches expand a time duration of 10 ns which is much smaller than the flat tops. Multi bunches will receive an uniform kick strength which insure a high injection efficiency as single bunch injection. The measure injection efficiency during top up operation is beyond 85%. One snapshot of injection efficiency is shown in Fig. 3.

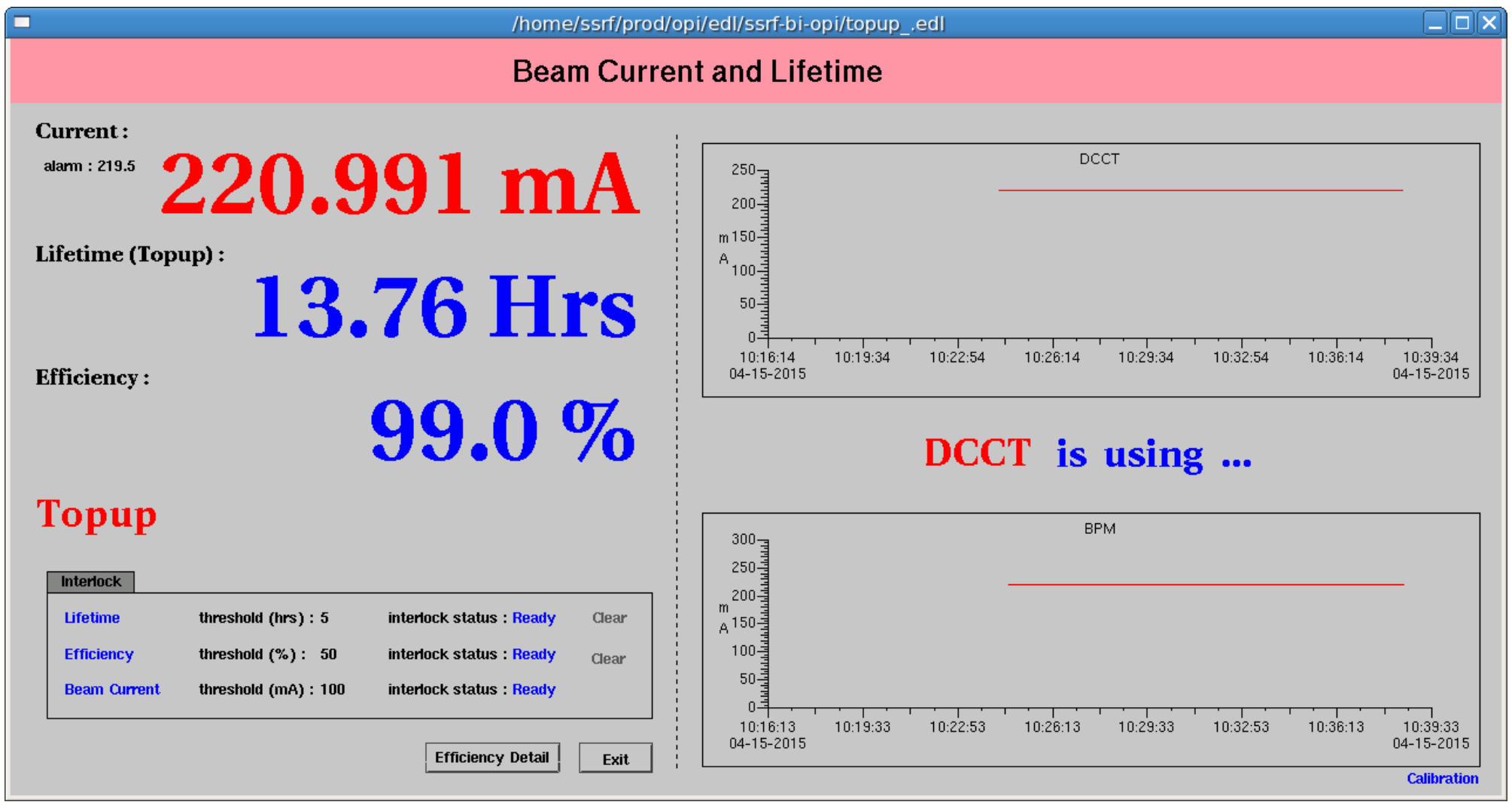

Fig 3. A snapshot of the machine status for top up operation.

To keep bunch charge uniform is much more difficult for the multi-bunch injection than the single bunch injection. For single bunch injection, the lowest charge bucket can be picked out by a bunch current monitor (BCM) and the top up injection controlling system will shot the injected bunch to the bucket. As for multi-bunch injection, the point to point refilling is impossible. Several methods have been tried to refill the current, including random shooting, injection bunches to buckets in a sequence and pick out lowest charge bucket to refill. Among them, the last one is the best for charge uniformity for top up operation. The injected charge distribution of the multi bunches is as shown in Fig. 4. It is important to shift the top up timing controller a 2 ns later to inject the highest charged bunch to the bucket holds lowest charge.

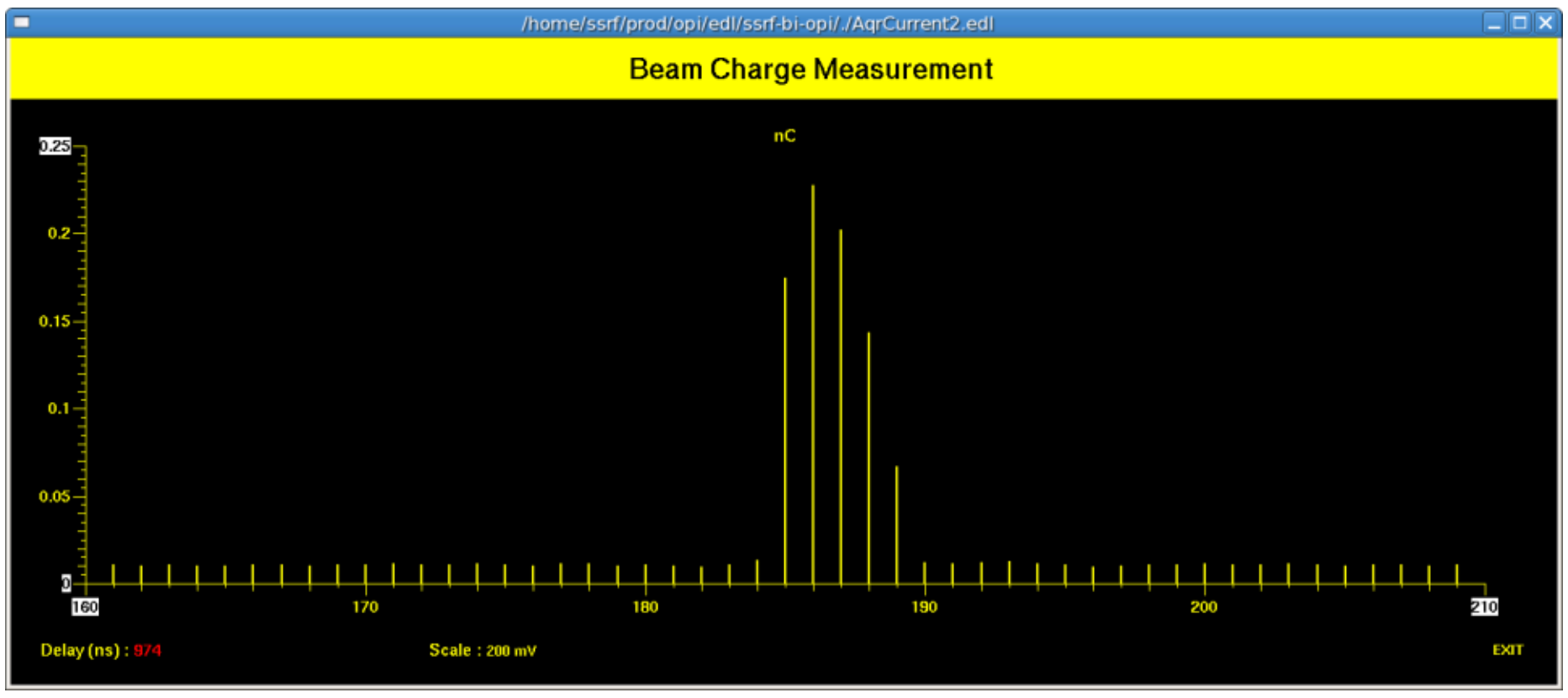

Figure 4. Injected multi-bunch charge distribution

The bunch charge deviation for multi-bunch injection is 3.3% as shown in Fig. 5a, deriving this number, the bunches at the two ends of the bunch trains are ignored since it is uncontrollable for multi-bunch injection and will not greatly effect BPMs. As a comparison, the bunch charge deviation for single bunch injection is 5.0% [9] as shown in Fig. 5b.

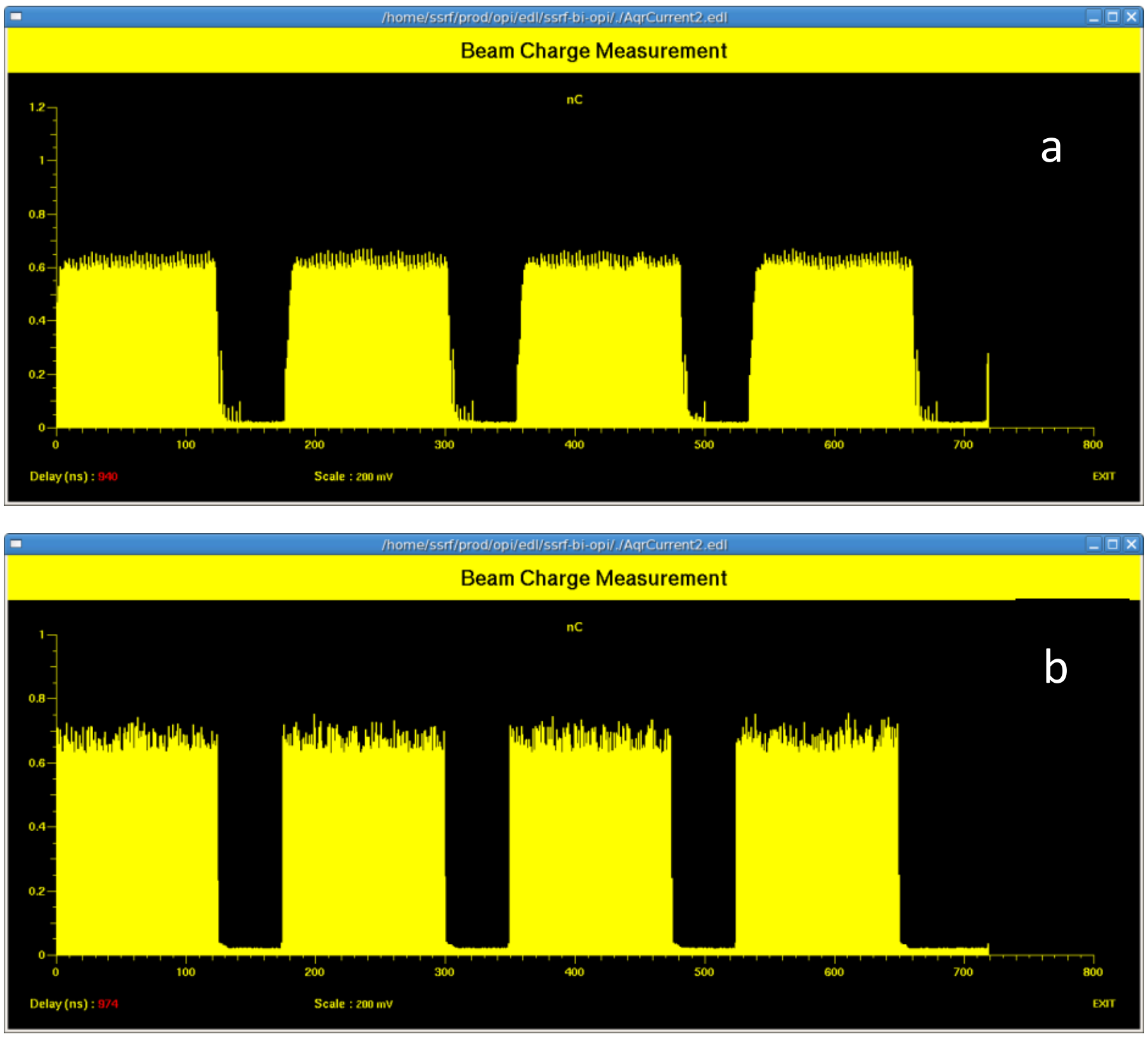

Figure 5. Bunch charge deviation during top up operation (Filling pattern 4*125bunch-train, total bucket number 720).

For multi-bunch injection, the injection rate can be more than 1.6 mA/second as shown in Fig. 6 which is much faster than single bunch injection rate 0.3 mA/second. The filling time of 240mA can be shortened from 20 minutes to 5 minutes. For top up refilling, single bunch injection requires 15 seconds while multi-bunch injection takes 4 seconds (For top up injection, the bunch charge is decreased by the focus solenoid after RF electron gun). Since for the local closed bump injection orbit distortion is inevitable, shorter the injection time better for users' experiments.

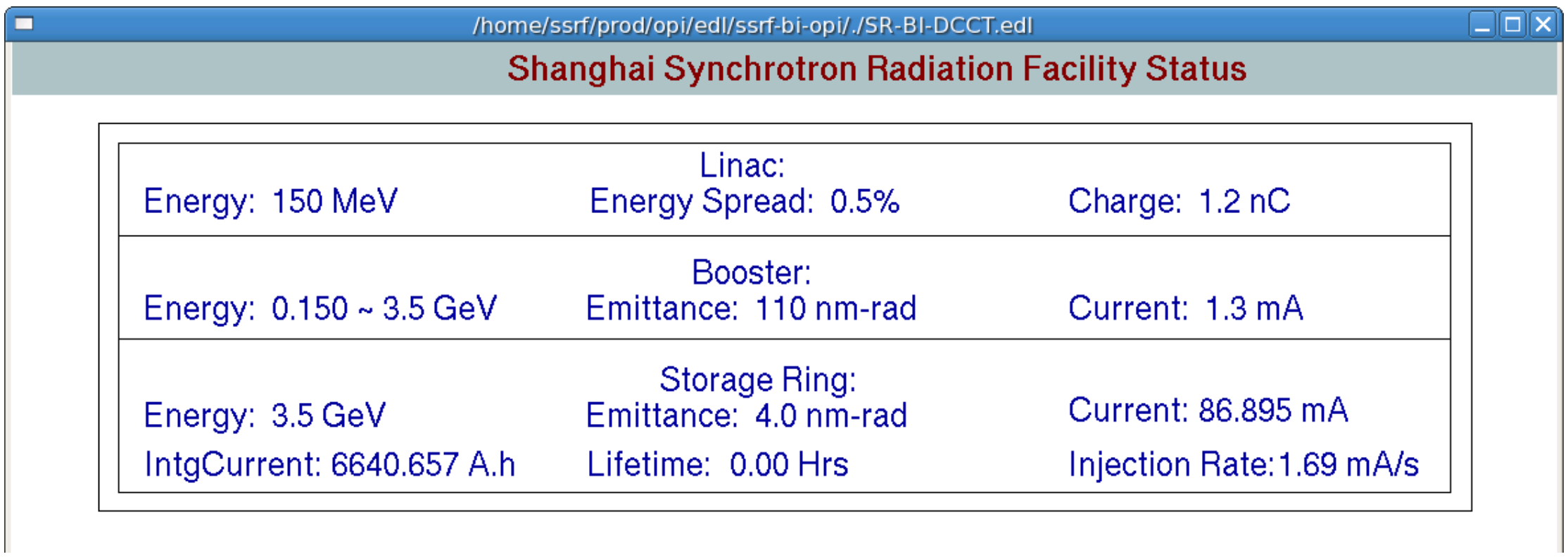

Figure 6. Injection rate of multi-bunch injection

## 5、Orbit stability

BPM readings are sensitive to the beam current and to the filling patterns. This is one of the main reasons for light sources to pursue top up operation to keep beam current and filling pattern constant providing a stable light for users. As has been shown in the preceding section the filling pattern is well kept constant during top up operation, the BPM noise from filling pattern can be minimized.

The orbit feedback systems for SSRF are global ones based on SVD algorism [10]. For SSRF, not all of the SVD singular values are used which makes the systems stable. The BPM reading errors from filling pattern change is not a real orbit distortion and can be filtered by SVD algorism. The measured 24 hours of orbit stability at the ends of straight section is as shown in Fig. 7, the RMS value for horizontal and vertical plane are 0.35μm and 0.13μm which is at the same level for single bunch injection (0.26 μm and 0.25μm).

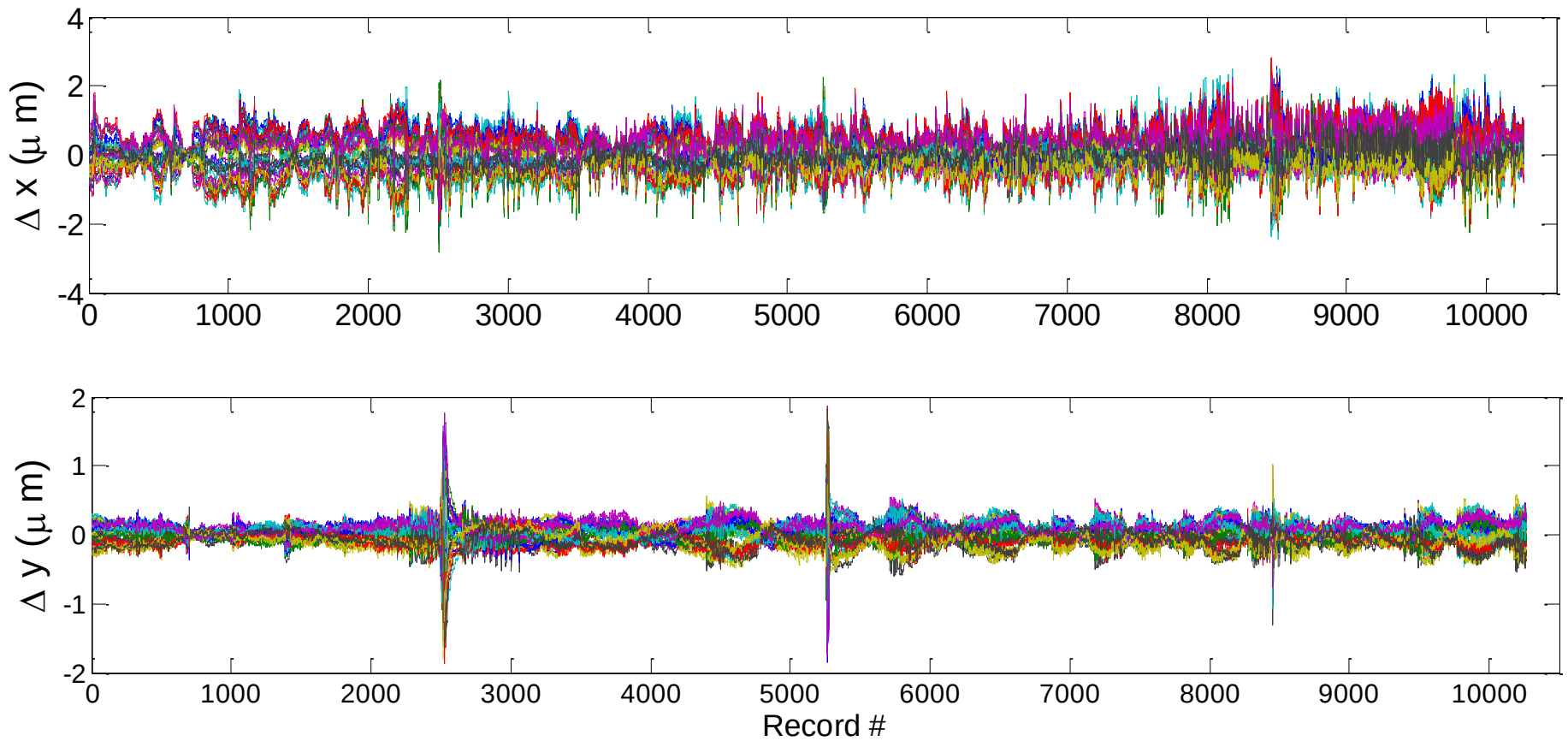

Figure 7. Orbit stability for multi-bunch top up injection.

## 6、Conclusions

The multi-bunch injection applied at SSRF had greatly shortened the injection time which will benefit shortening MDT and orbit distortion period for top up operation. The lifetime of pulsed injection components such as the septum, kickers are related to trigger frequency, it can be concluded that lifetime of the components will be prolonged.

Orbit stability is important for light sources. Filling pattern should be carefully arranged for multi-bunch injection to keep the bunch charge uniformly, minimizing the BPM noises.

As most of the performance of multi-bunch injection is similar to or better than single bunch injection, the scheme has been conducted for users' operation.

## Acknowledgments

This work was supported by National Natural Science Foundation of China under Grant no.11105214.

## Reference

[1] A. Olmos, F. Pérez, MEASUREMENTS ON LIBERA ELECTRON AND LIBERA BRILLIANCE BPM ELECTRONICS, Proceedings of BIW08, Tahoe City, California 2008, 194-196,.

[2] C. Christou, et al., THE PRE-INJECTOR LINAC FOR THE DIAMOND LIGHT SOURCE, Proceedings of LINAC 2004, Lübeck, Germany, 2004, 84-86.

[3] A. Setty, et al., COMMISSIONING OF THE 100 MEV PREINJECTOR HELIOS FOR THE SOLEIL SYNCHROTRON, Proceedings of EPAC 2006, Edinburgh, Scotland, 2006, 1274-1276.

[4] Sanghyun Park and Jeff Corbett, BOOSTER SYNCHROTRON RF SYSTEM UPGRADE FOR SPEAR3, Proceedings of IPAC'10, Kyoto, Japan, 2010, 2660-2662.

[5] J. M. Byrd and S. De Santis, PHYSICAL REVIEW SPECIAL TOPICS - ACCELERATORS AND BEAMS, VOLUME 4, 024401 (2001)

[6] C. Gough, et al., THE SLS BOOSTER SYNCHROTRON, Proceedings of EPAC 1998, Stockholm, Sweden, 1998, 584-586.

[7] K. Tamura, et al., SINGLE BUNCH PURITY DURING SPRING-8 STORAGE RING TOP-UP OPERATION, Proceedings of the 1st Annual Meeting of Particle Accelerator Society of Japan and the 29th Linear Accelerator Meeting in Japan, Funabashi Japan, 2004, 581-583.

[8] L. Emery, M. Borland, TOP-UP OPERATION EXPERIENCE AT THE ADVANCED PHOTON SOURCE, Proceedings of the 1999 Particle Accelerator Conference, New York, 1999, 200-202.

[9] Z.T.Zhao, et al., PROGRESS TOWARDS TOP-UP OPERATION AT SSRF, Proceedings of IPAC2011, San Sebastián, Spain, 2011, 3008-3010.

[10] B.C. Jiang, et al., STATUS OF THE SSRF FAST ORBIT FEEDBACK SYSTEM, Proceedings of IPAC2012, New Orleans, Louisiana, USA, 2012, 2855-2857.

# 上海光源储存环实现多脉冲注入

姜伯承，林国强， 汪宝亮，张满洲，殷重先，阎映炳，田顺强，王坤
中国科学院上海应用物理研究所，上海， 201800

*摘要*

上海光源采用了多束团注入，和单束团注入相比，它大大提高了注入速率、减小了注入时间。多束团注入将显著的降低用户运行时段的平均故障恢复时间同时提高脉冲注入原件的使用寿命。本文将详细介绍直线获得多束团的方法、注入填充效果及轨道稳定性的变化。